# Dipole Polarizability of Alkali-Metal (Na, K, Rb) – Alkaline-Earth-Metal (Ca, Sr) Polar Molecules: Prospects for Alignment

Geetha Gopakumar,[1,2]* Minori Abe,[1,2] Masahiko Hada,[1,2] and Masatoshi Kajita[3]

[1] *Department of Chemistry, Tokyo Metropolitan University, 1-1 Minami-Osawa, Hachioji, Tokyo 192-0397, Japan*

[2] *JST, CREST, 4-1-8 Honcho, Kawaguchi, Saitama 332-0012, Japan*

[3] *National Institute of Information and Communications Technology, Koganei, Tokyo 184-8795, Japan*

*geetha@tmu.ac.jp

## ABSTRACT

Electronic open-shell ground-state properties of selected alkali-metal (AM) – alkaline-earth-metal (AEM) polar molecules are investigated. We determine potential energy curves of the $^2\Sigma^+$ ground state at the coupled-cluster singles and doubles with partial triples (CCSD(T)) level of electron correlation. Calculated spectroscopic constants for the isotopes ($^{23}$Na, $^{39}$K, $^{85}$Rb) – ($^{40}$Ca, $^{88}$Sr) are compared with available theoretical and experimental results. The variation of the permanent dipole moment (PDM), average dipole polarizability, and polarizability anisotropy with internuclear distance is determined using finite-field perturbation theory at the CCSD(T) level. Owing to moderate PDM (KCa: 1.67 D, RbCa: 1.75 D, KSr: 1.27 D, RbSr: 1.41 D) and large polarizability anisotropy (KCa: 566 a.u., RbCa: 604 a.u., KSr: 574 a.u., RbSr: 615 a.u.), KCa, RbCa, KSr, and RbSr are potential candidates for alignment and orientation in combined intense laser and external static electric fields.

**Introduction**

Ultracold diatomic molecules are produced by binding two laser-cooled atoms via photoassociation[1] or by Feshbach resonance.[2] Currently, such molecules are produced from alkali-metal (AM) or alkaline-earth-metal (AEM) atoms. Taking advantage of the zero nuclear spin of even isotopes of AEM atoms, a better choice to create ultracold diatomic molecules would be to combine an even isotope AEM atom with itself or a different even isotope AEM atom. However, the binding energy of AEM–AEM diatomic molecules in the electronic ground state is generally shallow and the energy separations between different vibrational-rotational states are very narrow, which makes the localization in a single quantum state difficult.[3,4] In this context, it is appropriate to look for diatomic molecules produced by pairing even isotope AEM atoms with AM atoms for future precise experiments.

LiX (X: even isotope AEM atom) molecules seem to be the easiest to control in the internal state because Li is the lightest atom that can be laser cooled in the electronic ground state and the vibrational-rotational transition frequencies are the highest among the ultracold molecules. Therefore, LiX molecules are advantageous to observe the quantum degeneracy localizing all produced molecules in a single quantum state (including hyperfine). The energy structures of LiX molecules have been theoretically analyzed.[5–7] Using these results, it was also shown that the vibrational transition frequencies of optically trapped LiX molecules can be measured with uncertainties less than $10^{-16}$, which makes it possible to test the variation in the proton-to-electron mass ratio.[8–10] Groups at Kyoto University and University of Washington have succeeded in obtaining simultaneous degeneracy of Yb and Li atoms.[11,12] The Fourier transform spectrum was obtained with the LiCa molecule in a thermal beam.[13]

Production of LiX molecules is possible with both photoassociation and Feshbach resonance. Although production by Feshbach resonance is more advantageous to obtain densities of molecules in a single quantum state, it is experimentally challenging owing to the weak hyperfine interaction of the Li atom (hyperfine splitting is 152 MHz with $^6$Li and 402 MHz with $^7$Li) and the Feshbach resonance area being very narrow.[14] Therefore, it is useful to analyze other possible candidates using heavier AM atoms (Na, K, and Rb) instead of Li for the production of $^{23}$NaX, $^{39}$KX, and $^{85}$RbX molecules for ultracold experiments via photoassociation or Feshbach resonance. Among the AM atoms chosen, the $^{23}$Na atom is the second lightest AM atom that can be laser cooled. In addition, the amount of other isotopes is negligible, which makes the experiment easier. The hyperfine splitting of the $^{23}$Na atom is much larger than the Li atom (1.77 GHz), and hence the production of $^{23}$NaX molecules by Feshbach resonance is expected to be easier than LiX molecules. In comparison, the $^{85}$Rb atom has an even larger splitting (3.0 GHz), and hence production of $^{85}$RbX molecules by Feshbach resonance should be easier than both LiX and NaX molecules. However, for RbX and KX molecules, the energy gaps between different rotational states are small, and the permanent dipole moments are large. Therefore, these molecules are not useful for precise measurement of vibrational-rotational transition frequencies, but they are useful to study the long-range electric dipole–dipole interactions.

In this paper, we consider $^{23}$NaX, $^{39}$KX, and $^{85}$RbX (X = $^{40}$Ca and $^{88}$Sr) molecules similar to our previous study on $^6$LiX (X = $^{40}$Ca and $^{88}$Sr),[6,7] because both Ca and Sr atoms have already been cooled to ultralow temperatures. The outline of the paper is as follows. First, we present results of the potential energy curves (PECs) for the ground $^2\Sigma^+$ state using the coupled-cluster singles and doubles with partial triples (CCSD(T)) method with relativistic correlation-consistent atomic natural orbital (ANO-RCC)[15] basis sets. CCSD(T) is best suited for AM–AEM types of molecules because the ground state can be written with a single configuration over the entire potential curve. The electronic ground-state configuration of Na($^2S_{1/2}$) is [Ne]$(3s)^1(3p)^0$, K($^2S_{1/2}$) is [Ar]$(4s)^1(4p)^0$, Rb($^2S_{1/2}$) is [Kr]$(5s)^1(5p)^0$, Ca($^1S_0$) is [Ar]$(4s)^2(4p)^0$, and Sr($^1S_0$) is [Kr]$(5s)^2(5p)^0$. To determine the calculation accuracy, we compare the ground state spectroscopic constants with available theoretical and experimental values.

Electronic permanent dipole moment (PDM, $d$), and parallel ($\alpha_\parallel$) and perpendicular ($\alpha_\perp$) dipole polarizability components are obtained using finite-field perturbation theory (FFPT) with CCSD energy calculations. These values are further used in the determination of the average polarizability ($\bar{\alpha}$) and polarizability anisotropy ($\gamma$), which play an important role in aligning molecules in intense laser and static electric fields. In addition, considering a simple ellipsoidal charge distribution similar to the AM–AM molecule in Deiglmayr et al.,[16] the variation of the parallel and perpendicular components of the dipole polarizability with internuclear distance are also investigated. Finally, to check the accuracy of the calculation, we compare the 100 a.u. molecular calculations with the respective sums of atomic dipole polarizability.

## Methods and computational details

For all the calculations in this study, the MOLCAS code (version 7.2)[17] was used. Scalar relativistic effects were accounted for through the third-order Douglas–Kroll–Hess (DKH)[18,19] transformation of the relativistic Hamiltonian. We assumed $C_{2v}$ point-group symmetry for the energy calculations. We obtained potential energy curves and the corresponding spectroscopic constants at the spin-free level, considering the fact that spin-orbit effects are negligible for AM–AEM systems, as demonstrated with the LiYb molecule.[20]

The electronic ground states of the AM–AEM molecules were obtained using the CCSD(T) method. In CCSD(T) calculations, all of the core electrons below Na(1$s$), K(2$p$), Rb(3$d$), Ca(3$s$), and Sr(4$s$) including the ones mentioned are frozen and excitations are taken only from Na 2$s$, 2$p$, 3$s$; K 3$s$, 3$p$, 4$s$; Rb 4$s$, 4$p$, 5$s$; Ca 3$p$, 4$s$; and Sr 4$p$, 5$s$ orbitals. The ANO-RCC[15] basis sets used in this calculation are shown in Table 1. The basis set superposition error was assumed to be negligible because of the large number of basis functions.

We obtained the electronic PDM function and the static dipole polarizability using FFPT and taking the dipole field strength to be in the range of $\pm 10^{-4}$ a.u. to $\pm 10^{-5}$ a.u. followed by numerical derivative analysis. For the dipole polarizability calculations, the energy calculations were performed with $C_1$ point group symmetry. Variation of the PDM and static dipole polarizability function was first studied with respect to internuclear distance in the $x$, $y$, and $z$ directions. Using the three dipole polarizability components, the parallel component $\alpha_\parallel = \alpha_{zz}$ and perpendicular component $\alpha_\perp = \alpha_{xx}$ or $\alpha_{yy}$ were firstly obtained. We obtained the average polarizability $\bar{\alpha}$ and polarizability anisotropy ($\gamma$) using

$$\bar{\alpha} = (\alpha_\parallel + 2\alpha_\perp)/3 \qquad (1)$$

and

$$\gamma = \alpha_\parallel - \alpha_\perp. \qquad (2)$$

Considering the dipole polarizability of a charge distribution to be proportional to its volume, we investigated the relationships between the parallel and perpendicular dipole polarizabilities at the equilibrium bond distance ($R_e$) and the volume ($4\pi R_e^3/3$). In addition, the trends in the PDM, average polarizability, and polarizability anisotropy were compared with extensively available data on AM–AM systems.[16]

The equilibrium bond distance ($R_e$), harmonic frequency ($\omega_e$), rotational constant ($B_e$), dissociation energy ($D_e$), and PDM ($d_e$) were calculated using the VIBROT program in MOLCAS. Spectroscopic constants, vibrational PDM ($d_v$), vibrational average polarizability ($\bar{\alpha}_v$), and vibrational polarizability anisotropy ($\gamma_v$) were calculated for the range $R = 3$ to $30$ a.u. with 1000 grid points. The vibrational data were further used to define two dimensionless parameters proposed by Friedrich and Herscbach:[21–23]

$$\Delta\omega_{al} = \frac{\gamma_v I_L}{2B_v} \quad \text{for alignment} \qquad (3)$$

and

$$\omega_{or} = \frac{d_v \varepsilon_S}{B_v} \quad \text{for orientation.} \qquad (4).$$

Here, $I_L$ and $\varepsilon_S$ represents the intensity of the laser and the amplitude of the external static electric field. The values of these parameters can be conveniently evaluated in practical units as $\omega_{or} = 0.0168\, d_v\, \text{(Debye)}\varepsilon_S\, \text{(KV/cm)}/B_v(\text{cm}^{-1})$ and $\Delta\omega_{al} = 10^{-11}\, \gamma_v\, (\text{Å}^3)I_L\, (\text{W/cm}^{-2})/B_v(\text{cm}^{-1})$.

**Results and discussion**

**Potential energy curves and spectroscopic constants**

Figure 1(a) shows the potential energy curves obtained at the CCSD(T) level of electron correlation. These curves are drawn relative to the dissociation limit (AM ($^2S_{1/2}$) + AEM ($^1S_0$)) of each AM–AEM species, which has been calculated at the bond distance of 100.0 a.u. In Table 2, we show the ground-state spectroscopic constants for the isotopes ($^{23}$Na, $^{39}$K, $^{85}$Rb) – ($^{40}$Ca, $^{88}$Sr) using ANO-RCC basis sets at the CCSD(T) level. To the best of our knowledge, experimental as well as theoretical spectroscopic constants are not available in the literature for the NaCa, KCa, and RbCa systems. Hence, to determine the calculation accuracy, we compared the present calculation trend with our previous calculations of the LiCa molecule.[6,7] The spectroscopic constants of LiCa ($R_e$ = 3.395 Å, $\omega_e$ = 196.7 cm$^{-1}$, and $D_e$ = 2258 cm$^{-1}$) at CCSD(T) show the bond length decreases and $\omega_e$ and $D_e$ increase from the RbCa to the LiCa molecule. For AM – Sr molecules, in addition to our CCSD(T) estimates, the *ab initio* configuration interaction by perturbation selected iteration (CIPSI) calculations of Guerout *et al.*,[24] are also shown in Table 2. The CIPSI method gives a shorter bond length (~0.1 Å) and larger dissociation energy (~170 cm$^{-1}$) compared with our present CCSD(T) calculations. The trend in the CIPSI method (shorter bond length and larger dissociation energy) with respect to our CCSD(T) calculations was also noted in our earlier Li–AEM results[6]. The experimental spectroscopic constants ($R_e$ = 4.689 Å, and $D_e$ = 1000 cm$^{-1}$)[25] are only available for the RbSr molecule. As RbSr is the heaviest molecule studied, we estimated the errors in our present calculations on all AM–AEM polar molecules to be less than 1% and 10% for the bond lengths and dissociation energies, respectively.

In Figure 1(a), we have also included LiCa and LiSr PECs obtained at the same level of theory from our earlier calculations[6] for comparison. The PECs in Figure 1(a) only shift to a slightly larger internuclear distance when changing the AEM atom from Ca to Sr, whereas changing the AM atom (Li, Na, K, and Rb) clearly changes the PECs. Examining the results more closely, there is only a very small difference between KCa and RbCa (KSr and RbSr) molecules. This can be understood by analyzing the bond length of each molecule. The $R_e$ of the NaCa molecule is significantly shorter than that of the KCa and RbCa molecules, but significantly longer than the LiCa molecule. This result is reasonable considering that the bond length of AM–AEM molecules can be estimated by $r_a$(AM) + $r_a$(AEM), where $r_a$ denotes the atomic radius in picometers (1 Å = 100 pm) ($r_a$(Li) = 167 pm, $r_a$(Na) = 190 pm, $r_a$(K) = 243 pm, $r_a$(Rb) = 265 pm, $r_a$(Ca) = 194 pm, $r_a$(Sr) = 219 pm)[26].

Using the *ab initio* PECs, rovibrational spectra of the $^2\Sigma^+$ electronic ground states were also obtained. The spectroscopic constants of the $v$=0 vibrational state ($\omega_0$, $B_0$, and $D_0$) and the number of supported bound states for the angular momentum $J$=0 is also shown in Table 2.

**Permanent dipole moment, average dipole polarizability, and polarizability anisotropy**

Figure 1(b) shows the PDM as a function of internuclear distance $R$ at the CCSD(T) level of correlation using the FFPT method. The dipole field strengths were chosen after a few tests, taking care to reduce the discontinuities in the PDM functions. The dipole field strength was finally chosen to be similar to our earlier calculations of LiCa / LiSr systems ($\pm$ 0.00009 a.u.). In addition, we also checked the FFPT results with expectation values at the Hartree Fock (HF) level to determine the reliability of the chosen field strength.

We use the following sign convention for PDM values. A Positive PDM values indicate a charge transfer from the AM atom to the AEM atom, and a negative PDM values indicate the reverse. For all of the AM–AEM molecules, the PDM functions at the CCSD(T) level show similar behavior as a function of internuclear distance $R$, where the magnitude gradually increases as $R$ decreases, reaches a maximum at $R < R_e$, and then decreases. Theoretical calculations of NaSr, KSr, and RbSr by Guerout *et al*.,[24] using the CIPSI algorithm showed a similar trend to our calculations. The PDM functions in the region $R << R_e$ may not be reliable as the calculation showed instability when calculating the first derivatives at these positions. We obtained the PDM values at the equilibrium internuclear distances of 1.01 D for NaCa, 1.67 D for KCa, 1.75 D for RbCa, 0.49 D for NaSr, 1.27 D for KSr, and 1.41 D for RbSr. These values, particularly for KCa, RbCa, KSr, and RbSr, are relatively large, making them good candidates to study long-range dipole–dipole interactions. Theoretical calculations by Guerout *et al*.,[23] for NaSr, KSr and RbSr showed the PDM values at equilibrium internuclear distances to be 0.62 D, 1.52 D, and 1.54 D, which are all greater than the values from our FFPT calculations. Compared with the only available experimental PDM value for the RbSr molecule (1.36 D),[25] we infer that the PECs and PDMs using the FFPT method to be fairly accurate (~4%). Compared with well-studied AM–AM molecules such as KRb and RbCs, the PDMs of K/Rb–Ca/Sr molecules have values two times larger than the KRb molecule (0.6 D[27]) and similar to the RbCs molecule (1.2 D[28]). The ground rovibrational PDM ($d_v$) is also shown in Table 2.

The average dipole polarizability and polarizability anisotropy of the $^2\Sigma^+$ electronic ground state of the AM–AEM molecules are shown in Figure 2 and the values of the ground rovibrational levels are shown in Table 3. To the best of our knowledge, there are no theoretical or experimental values of the average polarizability and polarizability anisotropy for AM–AEM molecules. Hence, we compared the calculation trends with the available data of AM–AM molecules. First, we looked at the variation of the parallel and perpendicular components with internuclear distance. The $R$-variation of both components (Figure 3) showed a similar trend to AM–AM systems, with their magnitude increasing with increasing mass. This can also be clearly seen in Table 4, where the tabulated parallel and perpendicular dipole polarizability at the equilibrium internuclear distance shows an increasing trend when going from Li to Rb. In addition, for AM–AM systems, Deiglmayr et al.,[16] reported that the parallel polarizability component reaches a maximum at a distance around 1.3–1.5 times the equilibrium internuclear distance. Our calculations of AM–AEM systems, especially LiCa, LiSr, NaCa, and NaSr, shows a similar trend, except that the parallel polarizability component reaches a maximum at a shorter distance of around 1.2 times the $R_e$. The maximum parallel polarizability component of the KCa, RbCa, KSr, and RbSr molecules is exactly at $R_e$.

In contrast, the perpendicular components (Table 4) always have a smaller magnitude than the parallel component, and monotonically increase towards an asymptotic limit. As expected, at large distances ($R$ = 100 a.u., supermolecular limit), both components converge to the sum of the atomic components. We calculated the values of $\alpha_A$ by $\alpha_{AM} + \alpha_{AEM}$ using the experimental dipole polarizabilities of the atoms,[29] and the results are shown in Table 5. Our supermolecular data ($\alpha_{100}$) are close to the experimental values ($\alpha_A$) for all the molecules (well within the error bars of the experiments). Hence, we infer that the accuracy of the electric dipole polarizability calculations at the equilibrium internuclear regions to be of similar order.

To further investigate the trends, we plotted the parallel and perpendicular dipole polarizability components calculated at equilibrium internuclear distance as a function of volume of a sphere, $V_e = 4\pi R_e^3/3$. For AM–AM molecules, the components are aligned and a linear fit shows that the parallel component changes two times faster than the perpendicular component.[16] In Figure 4, we plot a similar graph for the AM–AEM molecules. Comparing the slopes of the straight line fits, the parallel component changes two times faster than the perpendicular component, as with AM–AM molecules.

Finally, we compare the magnitude of the average dipole polarizability and polarizability anisotropy of AM–AEM molecules with the extensively available literature of AM–AM molecules. For the AM–AM molecules (LiNa, LiK, LiRb, NaK, NaRb, and KRb), the average polarizability and polarizability anisotropy are in the range of 240–505 a.u. and 165–370 a.u., respectively. The AM–AEM molecules in the present study have larger average polarizability and polarizability anisotropy: 350–600 a.u. for the average polarizability and 228–615 a.u. for the polarizability anisotropy.

Furthermore, we also obtained the vibrationally averaged properties by solving the rovibrational Hamiltonian and taking the corresponding $R$ dependent properties at the CCSD(T) level of approximation. In Figure 5, the vibrationally averaged dipole polarizability and polarizability anisotropy of the ground electronic state are plotted against the vibrational index. Compared to the AM–AM molecules,[16] the LiCa, NaCa, LiSr, and NaSr molecules show a similar trend in which the vibrationally averaged dipole polarizability increases as a function of vibrational quantum number and eventually levels off for LiCa and LiSr. For LiCa, NaCa, LiSr and NaSr, the polarizability anisotropy reaches a maximum for $v$ in the range of 5-10, and then drops. In contrast, KCa, KSr, RbCa, and RbSr show a pattern of average polarizability increasing monotonically as $v$ increases, while the polarizability anisotropy monotonically decreases with $v$.

**Prospects of orientation and alignment of AM–AEM systems by combined laser and external electric fields**

The orientation and alignment of a molecule indicate a distribution of rotational levels with well-defined quantum number $M$ denoting the Z-component of the rotational angular momentum $J$. More specifically, orientation/alignment can be considered as the preferential direction/spatial distribution of $J$ with respect to the symmetry axis of the experiment.[30] It is well known that the PDM and dipole polarizability are important parameters to understand the possibility of confining the direction (orienting and aligning) of the molecular rotational axis in the presence of electric fields. To understand this further, we determined the two parameters ($\omega_{or}, \Delta\omega_{al}$) proposed by Friedrich and Herschbach[21–23] for aligning and orienting polar molecules by combining an intense laser field and an external electric field. To illustrate this new possibility offered by combined fields, Friedrich and Herschbach discussed typical values of $\omega_{or}$ and $\Delta\omega_{al}$ for different polar molecules (alkali halides, ICl, NO, and CO) using standard laser intensity and electric fields.[22] As an example, for the ICl molecule, $\omega_{or} = 5.5$ and $\Delta\omega_{al} = 780$ were obtained taking the standard laser intensity and electric fields to be $10^{12}$ W/cm$^2$ and 30 kV/cm, respectively.[22]

For a better comparison of the orientation and alignment properties of AM–AEM molecules with AM–AM molecules,[16] we have evaluated the magnitudes of the laser intensity and static electric field taking the Friedrich – Herschbach parameters $\omega_{or}$ and $\Delta\omega_{al}$ equal to one, respectively. These values give an indication of the field strengths required to achieve significant alignment or orientation of the molecules. The properties of the ground vibrational level of the X$^2\Sigma^+$ state that are relevant for orientation and alignment experiments are given in Table 6. For a comparison with AM–AM molecules, we also add data on RbCs (molecule with the largest anisotropy) and LiCs (molecule with the largest PDM) from Table VIII of Degleiyemer et al.[16]

Both AM–AEM and AM–AM molecules are found to be prospective candidates for alignment and orientation. Owing to their large anisotropies, both AM–AEM and AM–AM molecules can be easily aligned at lower intensity ($10^8$ W/cm$^2$) than for the ICl molecule. In addition, with moderate PDM and rotational constant, the static electric field required for permanent orientation is also one order of magnitude less (except for the LiCa, LiSr, and LiNa[16] molecules) than the ICl molecule. In summary, a large anisotropy makes AM–AEM molecules a better choice for alignment and a large PDM makes AM–AM molecules a better choice in situations requiring permanent orientation in combined fields. Let us investigate both situations more closely. Considering the RbSr molecule with the largest anisotropy ($\gamma_0$ = 612 a.u., $B_0$ = 0.017 cm$^{-1}$) with modest parameters for alignment ($\Delta\omega_{al}$ =1), the required intensity of the laser was found to be 1.9 × 10$^7$ W/cm$^2$. In a similar check of the RbCs molecule with the largest anisotropy among the AM–AM molecules ($\gamma_0$ = 441 a.u., $B_0$ = 0.029 cm$^{-1}$, $\Delta\omega_{al}$ = 1), the intensity of the laser required was found to be larger (4.4 × 10$^7$ W/cm$^2$),[16] but of similar order to that of the RbSr molecule. Furthermore, to check the suitability for experiments requiring orientation in external fields ($\omega_{or}$ = 1), we compared the molecules with the largest PDM between AM–AEM (RbCa: $d_0$ = 1.727 D, $B_0$ = 0.030 cm$^{-1}$) and AM–AM (LiCs: $d_0$ = −5.523 D, $B_0$ = 0.194 cm$^{-1}$) molecules. This shows the electric field to be 1.0 kV/cm for RbCa and 2.1 kV/cm for LiCs,[16] indicating that both molecules are good choices for further experiments.

Overall, KCa, RbCa, KSr, and RbSr are good candidates among the AM–AEM molecules for orientation and alignment with moderate laser intensities and external static fields owing to their large anisotropies and moderate PDMs. LiCa, LiSr, NaCa, and NaSr molecules have low anisotropies and PDM values and require larger laser power and static electric fields than the other AM–AEM molecules, although orientation and alignment are still experimentally possible.

**Conclusions**

In this paper, we have reported *ab initio* calculations of the electronic ground state ($^2\Sigma^+$) of AM–AEM (AM: Li, Na, K, Rb : AEM Ca, Sr,) systems. First, we investigated the potential energy curves and spectroscopic constants of the electronic ground state at the CCSD(T) level. We used the third-order Douglas–Kroll spin-free Hamiltonian and ANO-RCC basis sets. With the available theoretical and experimental spectroscopic constants, we estimated the errors in our present calculations to be less than 1% and 10% for bond lengths and dissociation energies, respectively.

Second, we obtained the electronic ground state PDM functions at the CCSD(T) level using the FFPT method. The obtained PDM values at the equilibrium internuclear distance were large, especially for KCa (1.67 D), RbCa (1.75 D), KSr (1.27 D), and RbSr (1.41 D) molecules. Compared with the available experimental and theoretical data, we estimate the errors to be around 4%. In addition, we also investigated the ground state average dipole polarizability and polarizability anisotropy over the internuclear distance using the FFPT method. We found very close agreement (within the experimental uncertainties) between our supermolecular polarizabilities and atomic polarizabilities for all the molecules. Our calculated data on average dipole polarizability and anisotropy provide a useful computational reference; to the best of our knowledge, no other reports currently exist. We also presented the vibrationally averaged properties, which are relevant for further investigation of these molecules. The complete sets of data (PECs, PDMs, and polarizabilities) are included in the supplementary material[31].

Finally, we investigated the conditions for orientation and alignment in AM–AEM molecules by combining intense laser and external electric fields. We found large anisotropies are advantageous for AM–AEM molecules, which could be further investigated in this field. We propose KCa, RbCa, KSr, and RbSr, which have moderate PDMs and large anisotropies, to be good candidates among the AM–AEM molecules considered in this study for future experiments involving alignment and orientation in the presence of combined laser and electric fields.

**Acknowledgments**

This research was supported by JST, CREST entitled "Creation of Innovative Functions of Intelligent Materials on the Basis of the Element Strategy".

**Keywords:** *ab initio*, ultracold molecules, dipole polarizability, anisotropy, alignment

**Table 1.** ANO-RCC basis set contraction style and the corresponding correlated orbitals at CCSD(T) level of theory.

| Elements | Contraction style | Correlated orbitals |
|:---:|:---:|:---:|
| Na | (14$s$9$p$4$d$3$f$1$g$)/[8$s$7$p$4$d$2$f$1$g$] | 2$s$, 2$p$, 3$s$ |
| K  | (17$s$12$p$6$d$2$f$)/[9$s$8$p$6$d$2$f$] | 3$s$, 3$p$, 4$s$ |
| Rb | (23$s$19$p$11$d$4$f$)/[10$s$10$p$5$d$4$f$] | 4s, 4$p$, 5$s$ |
| Ca | (20$s$16$p$6$d$4$f$)/[10$s$9$p$6$d$4$f$] | 3$p$, 4$s$ |
| Sr | (23$s$19$p$12$d$4$f$)/[11$s$10$p$7$d$4$f$] | 4$p$, 5$s$ |

**Table 2.** Spectroscopic constants [equilibrium distance ($R_e$), vibrational constant ($\omega_e$), rotational constant ($B_e$), dissociation energy ($D_e$), permanent dipole moment ($d_e$)] for ground $^2\Sigma^+$ states of $^{23}$Na$^{40}$Ca, $^{39}$K$^{40}$Ca, $^{85}$Rb$^{40}$Ca, $^{23}$Na$^{88}$Sr, $^{39}$K$^{88}$Sr, and $^{85}$Rb$^{88}$Sr at the CCSD(T) level of correlation. The vibrational spectroscopic constants of the $v = 0$ level are also shown along with the number of bound vibrational levels ($N_v$).

| Molecule | Method | $R_e$ (Å) | $\omega_e/\omega_0$ (cm$^{-1}$) | $B_e/B_0$ (cm$^{-1}$) | $D_e/D_0$ (cm$^{-1}$) | $N_v$ | $d_e$ (Debye) |
|---|---|---|---|---|---|---|---|
| NaCa | CCSD(T) | 3.72 | 97/92 | 0.083/0.083 | 1453/1407 | 31 | 1.01 |
| KCa | CCSD(T) | 4.32 | 61/58 | 0.045/0.045 | 974/944 | 32 | 1.67 |
| RbCa | CCSD(T) | 4.53 | 49/47 | 0.030/0.030 | 921/897 | 38 | 1.75 |
| NaSr | CCSD(T) | 3.89 | 82/79 | 0.061/0.061 | 1441/1401 | 35 | 0.49 |
|  | CIPSI[a] | 3.82 | 85 | 0.063 | 1597 |  | 0.62 |
| KSr | CCSD(T) | 4.53 | 48/47 | 0.030/0.030 | 964/940 | 38 | 1.27 |
|  | CIPSI[a] | 4.41 | 52 | 0.032 | 1166 |  | 1.52 |
| RbSr | CCSD(T) | 4.72 | 36/35 | 0.018/0.017 | 916/899 | 41 | 1.41 |
|  | CIPSI[a] | 4.60 | 32 | 0.018 | 1073 |  | 1.54 |
|  | Expt[b] | 4.69 |  |  | 1000 |  | 1.36 |

[a]Ref. [24], [b]Ref. [25]

**Table 3.** PDM values and polarizability values for the ground $^2\Sigma^+$ state of $^6$Li $^{40}$Ca, $^{23}$Na $^{40}$Ca, $^{39}$K $^{40}$Ca, $^{85}$Rb $^{40}$Ca, $^6$Li $^{88}$Sr, $^{23}$Na $^{88}$Sr, $^{39}$K $^{88}$Sr, and $^{85}$Rb $^{88}$Sr molecules calculated for the lowest vibrational level ($v = 0$).

| Molecule | $d_{v=0}$ (D) | $\bar{\alpha}_{v=0}$ (a.u.) | $\gamma_{v=0}$ (a.u.) |
|---|---|---|---|
| LiCa | 1.10 | 353 | 367 |
| NaCa | 1.00 | 354 | 227 |
| KCa | 1.64 | 515 | 562 |
| RbCa | 1.73 | 558 | 600 |
| LiSr | 0.31 | 399 | 377 |
| NaSr | 0.49 | 398 | 352 |
| KSr | 1.26 | 559 | 572 |
| RbSr | 1.39 | 599 | 612 |

**Table 4.** Polarizability values for the ground $^2\Sigma^+$ state of AM–AEM molecules calculated at the equilibrium internuclear distance.

| Molecule | $R_e$ (Å) | $\alpha_\parallel$ (a.u.) | $\alpha_\perp$ (a.u.) | $\bar{\alpha}$ (a.u.) | $\gamma$ (a.u.) |
|:---:|:---:|:---:|:---:|:---:|:---:|
| LiCa | 3.395 | 594 | 230 | 352 | 364 |
| NaCa | 3.716 | 581 | 240 | 354 | 228 |
| KCa | 4.324 | 892 | 326 | 515 | 566 |
| RbCa | 4.528 | 961 | 357 | 558 | 604 |
| LiSr | 3.531 | 621 | 271 | 395 | 372 |
| NaSr | 3.889 | 633 | 281 | 398 | 352 |
| KSr | 4.528 | 942 | 367 | 559 | 574 |
| RbSr | 4.724 | 1009 | 394 | 599 | 615 |

**Table 5.** Parallel, perpendicular, and average dipole polarizabilities obtained at the supermolecular limit (100.0 a.u.) for AM–AEM molecules at the CCSD(T) level of theory. Experimental and theoretical molecular dipole polarizabilities of the constituent atoms are compared with the dipole polarizability ($\alpha_{100}$) obtained at $R$ = 100.0 a.u. The atomic dipole polarizability ($\alpha_A$) is the sum of the atomic dipole polarizabilities of the of the two constituent atoms from Ref. [29].

| Molecule | $\alpha_\parallel$ (100) (a.u.) | $\alpha_\perp$ (100) (a.u.) | $\bar{\alpha}$(100) (a.u.) | $\alpha_A$ (a.u.) |
|---|---|---|---|---|
| LiCa | 328.2 | 328.1 | 328.1 | 333 ± 17 |
| NaCa | 320.9 | 320.8 | 320.8 | 331 ± 17 |
| KCa  | 449.8 | 449.5 | 449.6 | 460 ± 18 |
| RbCa | 492.4 | 491.9 | 492.1 | 488 ± 18 |
| LiSr | 370.4 | 370.3 | 370.3 | 350 ± 15 |
| NaSr | 362.9 | 362.8 | 362.8 | 349 ± 15 |
| KSr  | 489.0 | 488.7 | 488.8 | 477 ± 16 |
| RbSr | 538.9 | 538.6 | 538.7 | 505 ± 16 |

**Table 6.** Vibrational properties (polarization anisotropy ($\gamma_v$), rotational constant ($B_v$), and PDM ($d_v$)) of the $v$=0 level and the corresponding properties of the level with largest anisotropy of $^6$Li$^{40}$Ca, $^{23}$Na$^{40}$Ca, $^{39}$K$^{40}$Ca, $^{85}$Rb$^{40}$Ca, $^6$Li$^{88}$Sr, $^{23}$Na$^{88}$Sr, $^{39}$K$^{88}$Sr, and $^{85}$Rb$^{88}$Sr molecules at the CCSD(T) level of theory. The order of the parameters laser intensity ($I_L$), and the external static electric field ($\varepsilon_S$) are obtained by taking $\omega_{or}$ =1 and $\Delta\omega_{al}$ =1. For comparison with AM–AM molecules, the corresponding data for RbCs (largest anisotropy) and LiCs (largest PDM) molecules are included from Ref. [16]. Note: For KCa and RbCa molecules, the $v$ = 0 level has the largest anisotropy. For the AM–AM molecule (for eg. RbCs) [28], a negative PDM implies an excess electron charge on the Rb atom.

|      | $v$ | $\gamma_v$ (a.u.) | $B_v$ (cm$^{-1}$) | $d_v$ (D) | $I_L$ ($10^8$W/cm$^2$) | $\varepsilon_S$ (kV/cm) |
|------|----|-------|-------|--------|-------|-------|
| LiCa | 0  | 367   | 0.276 | 1.099  | 5.08  | 15.0  |
|      | 4  | 380   | 0.250 | 0.862  | 4.43  | 17.3  |
| NaCa | 0  | 227   | 0.083 | 0.995  | 2.46  | 5.0   |
|      | 5  | 232   | 0.075 | 0.787  | 2.18  | 5.7   |
| KCa  | 0  | 562   | 0.045 | 1.638  | 0.54  | 1.6   |
| RbCa | 0  | 600   | 0.030 | 1.727  | 0.34  | 1.0   |
| LiSr | 0  | 377   | 0.232 | 0.311  | 4.15  | 44.4  |
|      | 3  | 400   | 0.205 | 0.213  | 3.46  | 49.7  |
| NaSr | 0  | 352   | 0.061 | 0.488  | 1.17  | 7.4   |
|      | 5  | 369   | 0.056 | 0.423  | 1.02  | 7.9   |
| KSr  | 0  | 572   | 0.030 | 1.258  | 0.35  | 1.4   |
|      | 1  | 573   | 0.030 | 1.229  | 0.35  | 1.4   |
| RbSr | 0  | 612   | 0.017 | 1.395  | 0.19  | 0.7   |
|      | 1  | 612   | 0.017 | 1.369  | 0.19  | 0.8   |
| RbCs | 0  | 441   | 0.029 | −1.237 | 0.44  | 1.4   |
|      | 77 | 488   | 0.017 | −0.906 | 0.23  | 1.1   |
| LiCs | 0  | 327   | 0.194 | −5.523 | 4.0   | 2.1   |
|      | 41 | 565   | 0.108 | −3.051 | 1.3   | 2.1   |

**Figure 1.** (Color online) (a) Potential energy curves and (b) permanent dipole moments (PDM) of the ground $^2\Sigma^+$ state for AM–AEM molecules (AM: Li, Na, K, and Rb; AEM : Ca and Sr;) at the CCSD(T) level of correlation.

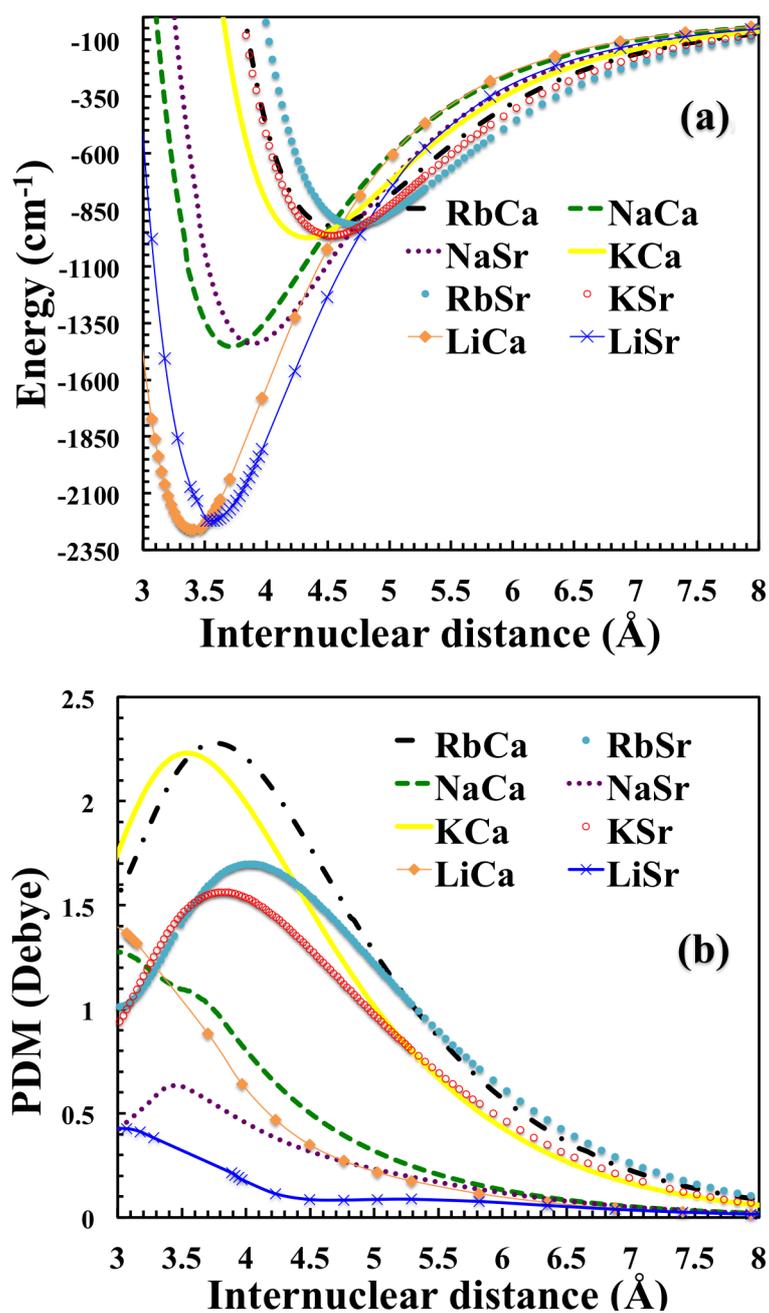

**Figure 2.** (Color online) (a) Average dipole polarizability and (b) polarizability anisotropy of the ground $^2\Sigma^+$ state for AM–AEM molecules (AM: Li, Na, K, and Rb; AEM: Ca and Sr;).

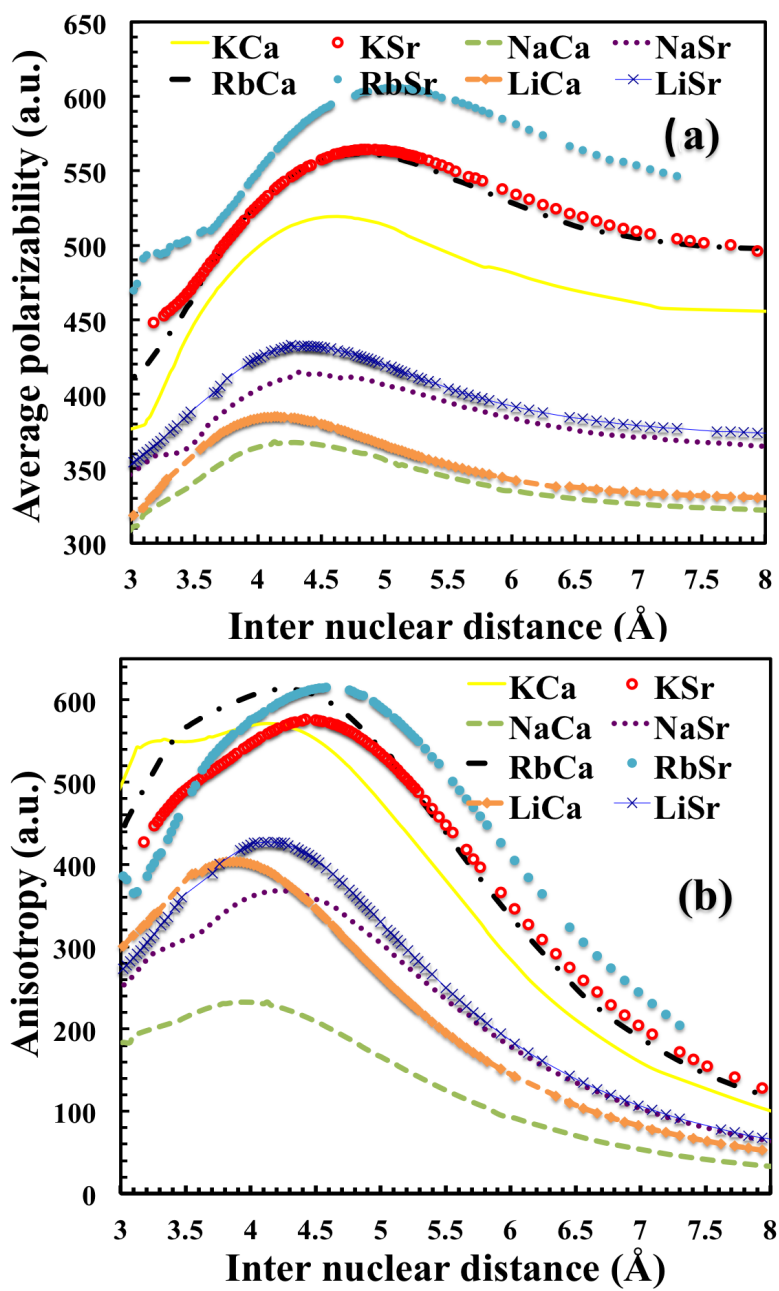

**Figure 3.** (Color online) (a) Parallel dipole polarizability and (b) perpendicular dipole polarizability of the ground $^2\Sigma^+$ state for AM–AEM molecules (AM: Li, Na, K, and Rb ; AEM: Ca and Sr).

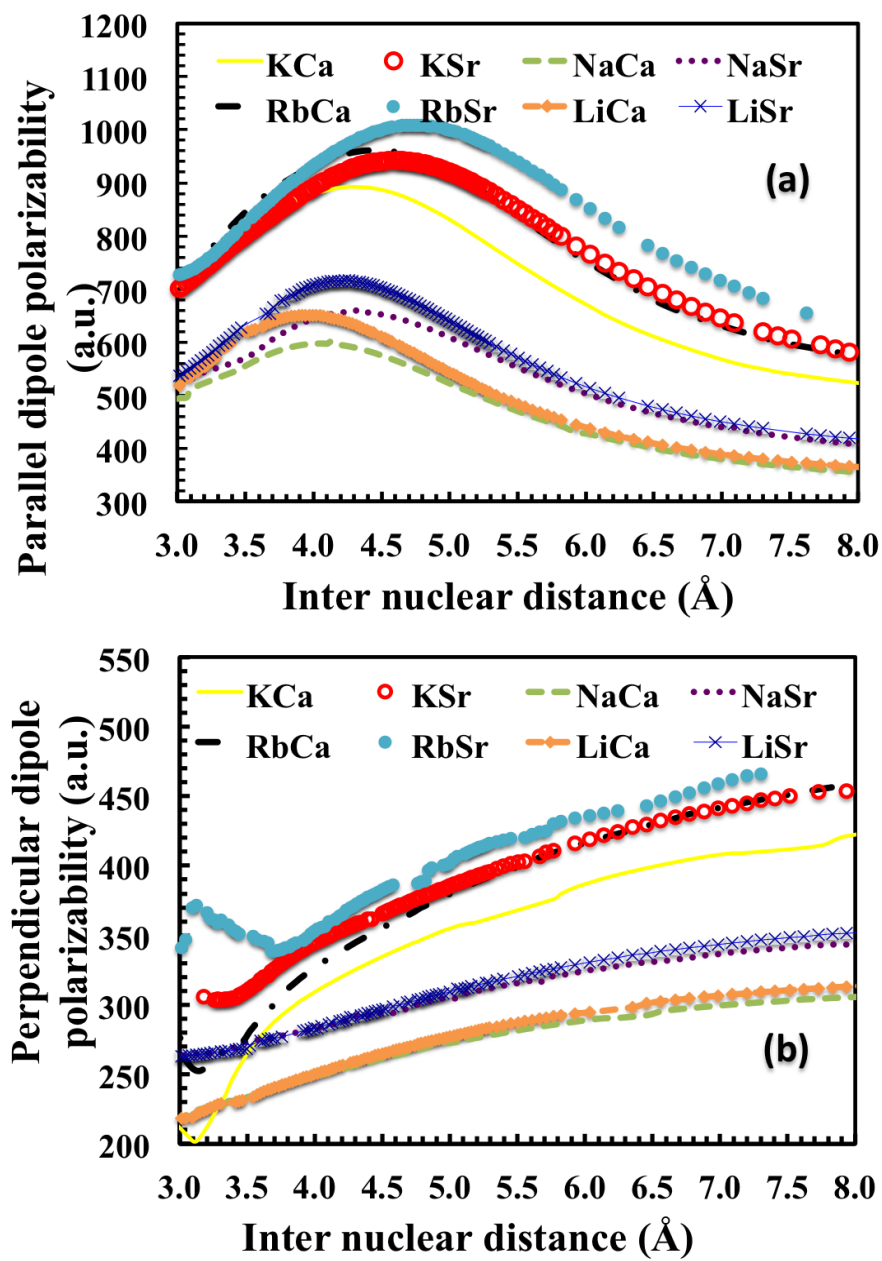

**Figure 4.** (Color online) Parallel and perpendicular dipole polarizability of the ground $^2\Sigma^+$ state for AEM–AM molecules (AM: Li, Na, K, and Rb ; AEM: Ca and Sr) as a function of $V_e = 4\pi R_e^3/3$, where $R_e$ is the equilibrium distance of the ground state of the AM–AEM pair. The straight lines are a linear fit of this variation in a.u. corresponding to $\alpha_\parallel = 0.254\ V_e + 272.05$ (dashed line) and $\alpha_\perp = 0.1396\ V_e + 55.318$ (solid line).

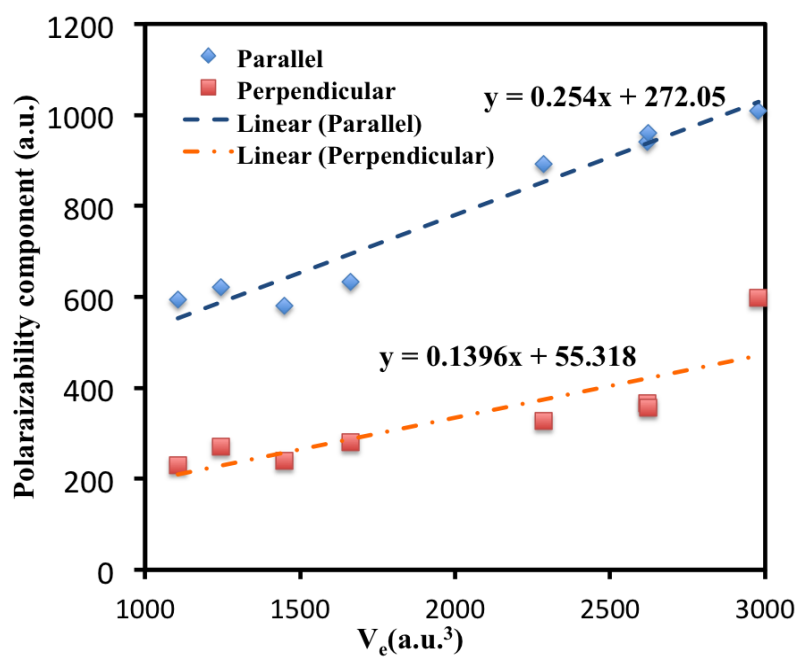

**Figure 5.** (Color online) (a) Vibrational average dipole polarizability and (b) vibrational polarizability anisotropy of the ground $^2\Sigma^+$ state for AM–AEM molecules (AM: Li, Na, K, and Rb ; AEM: Ca and Sr).

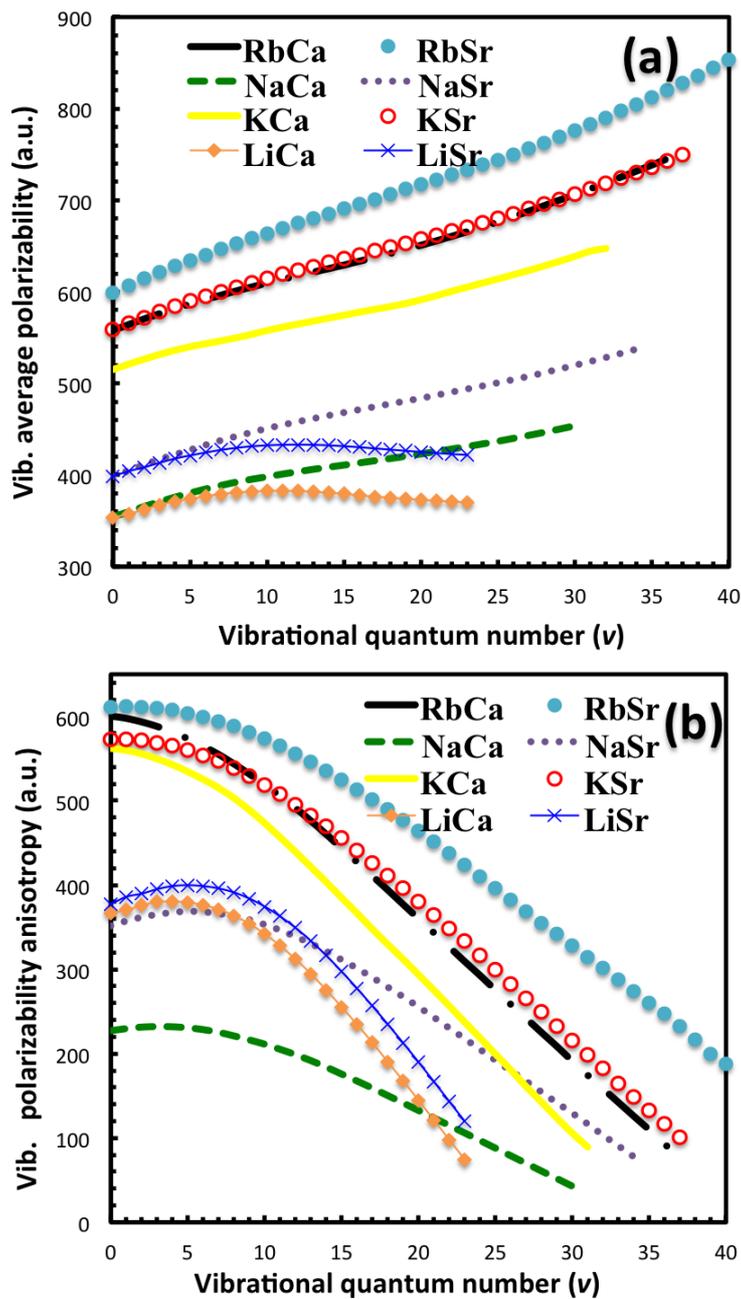